\documentclass[useAMS,usenatbib]{mn2e}
\usepackage{mn2e-breakabs}
\usepackage{graphicx}
\usepackage{times}
\usepackage{float}
\usepackage{hyperref}
\usepackage{subfig}
\usepackage{amsmath}
\def\la{\mathrel{\mathpalette\fun <}}
\def\ga{\mathrel{\mathpalette\fun >}}
\def\fun#1#2{\lower3.6pt\vbox{\baselineskip0pt\lineskip.9pt
        \ialign{$\mathsurround=0pt#1\hfill##\hfil$\crcr#2\crcr\sim\crcr}}}

\newcommand{\be}{\begin{equation}}
\newcommand{\ee}{\end{equation}}
\newcommand{\ba}{\begin{eqnarray}}
\newcommand{\ea}{\end{eqnarray}}
\newcommand{\simgt}{\,\hbox{\lower0.6ex\hbox{$\sim$}\llap{\raise0.6ex\hbox{$>$}}}\,}
\newcommand{\simlt}{\,\hbox{\lower0.6ex\hbox{$\sim$}\llap{\raise0.6ex\hbox{$<$}}}\,}

\begin{document}

\title
{More Evidence for the Redshift Dependence of Color from the JLA Supernova Sample Using Redshift Tomography}

\author[Miao Li et al.]
{
  \parbox{\textwidth}{
    Miao Li$^{1}$\thanks{mli@itp.ac.cn},
    Nan Li$^{1,2,3,4}$\thanks{linan@itp.ac.cn},
    Shuang Wang$^{1}$\thanks{wangshuang@mail.sysu.edu.cn (Corresponding author)}
    Zhou Lanjun$^{1,2,3,4}$\thanks{ljzhou@itp.ac.cn}
}
  \vspace*{4pt} \\
$^1$ School of Astronomy and Space Science, Sun Yat-Sen University, Guangzhou
510275, P. R. China\\
$^2$ Key Laboratory of Theoretical Physics, Institute of Theoretical Physics,
Chinese Academy of Sciences, Beijing 100190, P. R. China\\
$^3$ Kavli Institute for Theoretical Physics China,
Chinese Academy of Sciences, Beijing 100190, P. R. China\\
$^4$ School of Physical Science, University of Chinese Academy of Science, Beijing
100049, China\\
}

\date{\today}

\maketitle

\begin{abstract}

In this work, by applying the redshift tomography method to Joint Light-curve
Analysis (JLA) supernova sample, we explore the possible redshift-dependence of
stretch-luminosity parameter $\alpha$ and color-luminosity parameter $\beta$. The
basic idea is to divide the JLA sample into different redshift bins, assuming
that $\alpha$ and $\beta$ are piecewise constants. Then, by
constraining the $\Lambda$CDM model, we check the consistency of
cosmology-fit results given by the SN sample of each redshift bin.
We also adopt the same technique to explore the possible evolution of $\beta$
in various subsamples of JLA. Using the full JLA data, we find
that $\alpha$ is always consistent with a constant. In contrast, at high redshift
$\beta$ has a significant trend of decreasing, at $\sim 3.5\sigma$ confidence
level (CL). Moreover, we find that low-$z$ subsample favors a constant $\beta$;
in contrast, SDSS and SNLS subsamples favor a decreasing $\beta$ at 2$\sigma$ and
$3.3\sigma$ CL, respectively. Besides, by using a binned parameterization
of $\beta$, we study the impacts of $\beta$'s evolution on parameter estimation.
We find that compared with a constant $\beta$, a varying $\beta$ yields a larger
best-fit value of fractional matter density $\Omega_{m0}$, which slightly deviates
from the best-fit result given by other cosmological observations. However, for
both the varying $\beta$ and the constant $\beta$ cases, the $1\sigma$ regions of
$\Omega_{m0}$ are still consistent with the result given by other observations.

\end{abstract}


\section{Introduction}  \label{sec:intro}
Type Ia supernova (SN Ia) is a sub-category of cataclysmic variable stars that
results from the violent explosion of a white dwarf star in a binary system
\cite{Hillebrandt2000}.
It can be used as standard candles to measure the expansion history of the universe
\cite{Riess1998,Perl1999}, and it has become one of the most powerful tools to
probe the nature of dark energy (DE) \cite{Frieman08,Wang10,LLWW11,LLWW13,Weinberg13}.
In recent years, several supernova (SN) datasets have been released, such as
``SNLS''~\cite{Astier06}, ``Union''~\cite{Kowalski08}, ``Constitution'' ~\cite{Hicken09a,Hicken09b},
``SDSS''~\cite{Kessler09}, ``Union2''~\cite{Amanullah10}, ``SNLS3''~\cite{Conley2011} and ``Union2.1''~\cite{Suzuki12}.
The latest SN sample is ``Joint Light-curve Analysis'' (JLA) dataset \cite{Betoule2014},
which consists of 740 supernovae (SNe). JLA data includes 118 SNe at $0<z<0.1$
from several low-redshift samples~\citep{Hamuy1996, Riess1999, Jha2006, Contreras2010, Hicken09a, Hicken09b},
374 SNe at $0.03<z<0.4$ from the Sloan Digital Sky Survey (SDSS) SN search~\cite{Holtzman2008},
239 SNe at $0.1<z<1.1$ from the Supernova Legacy Survey (SNLS) observations~\cite{Guy2010}
and 9 SNe at $0.8<z<1.3$ from Hubble Space Telescope (HST)~\cite{Riess2007}.
It should be stressed that, in the process of cosmology-fits,
Betoule et al. treated two important quantities, stretch-luminosity parameter $\alpha$
and color-luminosity parameter $\beta$ of SN Ia, as free model parameters \cite{Betoule2014}.
This procedure is same as the recipe of \cite{Conley2011}.

The early proposals to use SN Ia as standard candles made an assumption that
the early samples were too small to test. By now SN samples are large enough for
many meaningful tests to be done. One of the most important tests is to probe the
possibility of redshift-dependence of $\alpha$ and $\beta$.
So far, there is no evidence for the evolution of $\alpha$.
But the redshift-dependence of $\beta$ has been found for several SN datasets.
For examples, by using the bin-by-bin method,
Marriner et al. found the redshift-dependence of $\beta$ for the SDSS data \cite{Marriner11}.
Besides, by adopting a linear $\beta$,
Mohlabeng and Ralston found the evolution of $\beta$ at 7$\sigma$ confidence level (CL)
for the Union2.1 data \cite{Mohlabeng13}.
In addition, one of the present authors had also done a series of research works about this issue.
In \cite{WangWang2013}, we found that $\beta$ deviates from a constant at 6$\sigma$ CL for the SNLS3 data.
Soon after, by studying various DE and modified gravity models with a linear $\beta$ \cite{WLZ14,WWGZ14,WWZ14,Wang15},
we found that the evolution of $\beta$ has significant effects on parameter estimation,
and the introduction of a time-varying $\beta$ can reduce the tension between SN Ia and other cosmological observations.

In a recent work \cite{Shariff15}, the discussion about time-varying $\beta$ has
been extended into the case of JLA data.
By adopting two specific parameterizations of $\beta$,
Shariff et al. found 4.6$\sigma$ CL evidence for a significant drop in $\beta$ at redshift $z=0.66$ \cite{Shariff15}.
It should be pointed out that, the results of \cite{Shariff15} depend on two
particular parameterizations of $\beta$.
To further investigate the possible redshift-dependence of $\beta$,
it is necessary to revisit this issue using a model-independent method.
In this work, we adopt the redshift tomography method,
which has been widely used in the investigation of cosmology \cite{Marriner11,Cai14,Giannantonio15}.
The basic idea is to divide the SN data into different redshift bins, assuming
that both $\alpha$ and $\beta$ are piecewise constants.
It should be pointed out that, adopting the redshift tomography method
will reduce the statistical significance.
Then we constrain $\Lambda$-cold-dark-matter ($\Lambda$CDM) model and check the consistency of
cosmology-fit results in each bin. In addition, it is very interesting to explore
the possible evolution of $\beta$ in various subsamples of JLA.
As far as we know, this issue has not been studied in the past. Therefore,
we also apply the same technique to various subsamples of JLA. Moreover, it is important
to study the impacts of possible redshift-dependence of $\beta$ on the parameter estimation.
To do this, we adopt a binned parameterization of $\beta$ in the analysis.

We describe our method in section \ref{sec:data}, present our results in section
\ref{sec:results}, and summarize in section \ref{sec:conclusion}.

\section{Methodology}
\label{sec:data}
In this section, we firstly introduce how to calculate the $\chi^2$ function of JLA data.
Then, we describe the details of the redshift tomography method.

Theoretically, the distance modulus $\mbox{\bf $\mu$}_{th}$ in a flat universe can be written as
\begin{equation}
  \mbox{\bf $\mu$}_{th} = 5 \log_{10}\bigg[\frac{d_L(z_{hel},z_{cmb})}{Mpc}\bigg] + 25,
\end{equation}
where $z_{cmb}$ and $z_{hel}$ are the CMB restframe and heliocentric redshifts of SN.
The luminosity distance ${d}_L$ is given by
\begin{equation}
  {d}_L(z_{hel},z_{cmb}) = \frac{(1+z_{hel})c}{H_0} \int_0^{z_{cmb}} \frac{dz}{E(z)},
\end{equation}
where $c$ is the speed of light, $H_0$ is the Hubble constant and $E(z) \equiv H(z)/H_0$
is the reduced Hubble parameter. For $\Lambda$CDM,
$E(z)$ can be written as
\begin{equation}
  E(z)=\sqrt{\Omega_{m0}(1+z)^3+(1-\Omega_{m0})}.
\end{equation}
Here $\Omega_{m0}$ is the present fractional matter density.

The observation of distance modulus $\mbox{\bf $\mu$}_{obs}$ is given by a
empirical linear relation:
\begin{equation}
  \mbox{\bf $\mu$}_{obs}= m_{B}^{\star} - M_B + \alpha \times X_1
  -\beta \times {\cal C},
\end{equation}
where $m_B^{\star}$ is the observed peak magnitude in the rest-frame
\text{of the} $B$ band,
$X_1$ describes the time stretching of light-curve, ${\cal C}$ describes the
supernova color at maximum brightness and $M_B$ is the absolute B-band magnitude,
which depends on the host galaxy properties \cite{Schlafly11,Johansson13}.
Notice that $M_B$ is related to the host stellar mass
($M_{stellar}$) by a simple step function \cite{Betoule2014}
\begin{equation}
  \label{eq:mabs}
    M_B = \left\lbrace
   \begin{array}{ll}
    M^1_B &\quad \text{if}\quad  M_\text{stellar} < 10^{10}~M_{\odot}\,,\\
    M^2_B &\quad \text{otherwise.}
    \end{array}
    \right.
\end{equation}
Here $M_{\odot}$ is the mass of sun.

The $\chi^2$ of JLA data can be calculated as
\begin{equation}
  \chi^2 = \Delta \mbox{\bf $\mu$}^T \cdot \mbox{\bf Cov}^{-1} \cdot \Delta\mbox{\bf $\mu$},
\end{equation}
where $\Delta \mbox{\bf $\mu$}\equiv \mbox{\bf $\mu$}_{obs}-\mbox{\bf $\mu$}_{th}$
is the data vector and $\mbox{\bf Cov}$ is the total covariance matrix, which is given
by
\begin{equation}
\mbox{\bf Cov}=\mbox{\bf D}_{\rm stat}+\mbox{\bf C}_{\rm stat}
+\mbox{\bf C}_{\rm sys}.
\end{equation}
Here $\mbox{\bf D}_{\rm stat}$ is the diagonal part of the statistical
uncertainty, which is given by \cite{Betoule2014},
\begin{eqnarray}
\mbox{\bf D}_{\rm stat,ii}&=&\left[\frac{5}{z_i \ln 10}\right]^2 \sigma^2_{z,i}+
  \sigma^2_{\rm int} +\sigma^2_{\rm lensing} + \sigma^2_{m_B,i} \nonumber\\
&&   +\alpha^2 \sigma^2_{X_1,i}+\beta^2 \sigma^2_{{\cal C},i}
+ 2 \alpha C_{m_B X_1,i} - 2 \beta C_{m_B {\cal C},i}\nonumber\\
&&  -2\alpha\beta C_{X_1 {\cal C},i},
\end{eqnarray}
where the first three terms account for the uncertainty in redshift due to peculiar velocities,
the intrinsic variation in SN magnitude and the variation of magnitudes caused by gravitational lensing.
$\sigma^2_{m_B,i}$, $\sigma^2_{X_1,i}$, and $\sigma^2_{{\cal C},i}$
denote the uncertainties of $m_B$, $X_1$ and ${\cal C}$ for the $i$-th SN.
In addition, $C_{m_B X_1,i}$, $C_{m_B {\cal C},i}$ and $C_{X_1 {\cal C},i}$
are the covariances between $m_B$, $X_1$ and ${\cal C}$ for the $i$-th SN.
Moreover, $\mbox{\bf C}_{\rm stat}$ and $\mbox{\bf C}_{\rm sys}$
are the statistical and the systematic covariance matrices, given by
\begin{equation}
\mbox{\bf C}_{\rm stat}+\mbox{\bf C}_{\rm sys}=V_0+\alpha^2 V_a + \beta^2 V_b +
2 \alpha V_{0a} -2 \beta V_{0b} - 2 \alpha\beta V_{ab},
\end{equation}
where $V_0$, $V_{a}$, $V_{b}$, $V_{0a}$, $V_{0b}$ and $V_{ab}$ are matrices given
by the JLA group at the link:
http://supernovae.in2p3.fr/sdss$_{-}$snls$_{-}$jla/ReadMe.html.
For the detailed discussions about JLA SN sample, see Ref. \cite{Betoule2014}.

As pointed out in \cite{Betoule2014}, in the process of calculating $\chi^2$,
both the Hubble constant $H_0$ and the absolute B-band magnitude $M_B$ are marginalized.
In this work, we follow the procedure of \cite{Betoule2014}, and do not treat $H_0$ and $M_B$ as free parameters.
We refer the reader to Ref. \cite{Betoule2014}, as well as the code of the JLA likelihood for the details of calculation.

As mentioned above, our aim is to explore the possible evolution of SN using a
model-independent method. In this work, we adopt the redshift tomography method.
The basic idea is to divide the SN sample into different redshift bins,
assuming that both $\alpha$ and $\beta$ are are piecewise constants.
Then, by constraining the $\Lambda$CDM model, we check the consistency of
cosmology-fit results given by the SN sample of each redshift bin. Moreover,
to ensure that our results are insensitive to the details of redshift
tomography, we evenly divide the JLA sample at redshift region [0,1] into 3 bins,
4 bins and 5 bins, respectively; then, we compare the fitting results obtained from these three
cases. In this work we perform a MCMC likelihood analysis using the ``CosmoMC''
package \citep{Lewis02}.

\section{Result}
\label{sec:results}

In this section, we mainly focus on the evolution behaviors of luminosity
standardization parameters $\alpha$ and $\beta$. Firstly, we present the results
given by the full JLA sample; then, we present the results given by various
subsamples of JLA; finally, we discuss the impacts of time-varying $\beta$ on
parameter estimation.

\begin{figure*}
    \centering
      \resizebox{0.9\columnwidth}{!}{\includegraphics{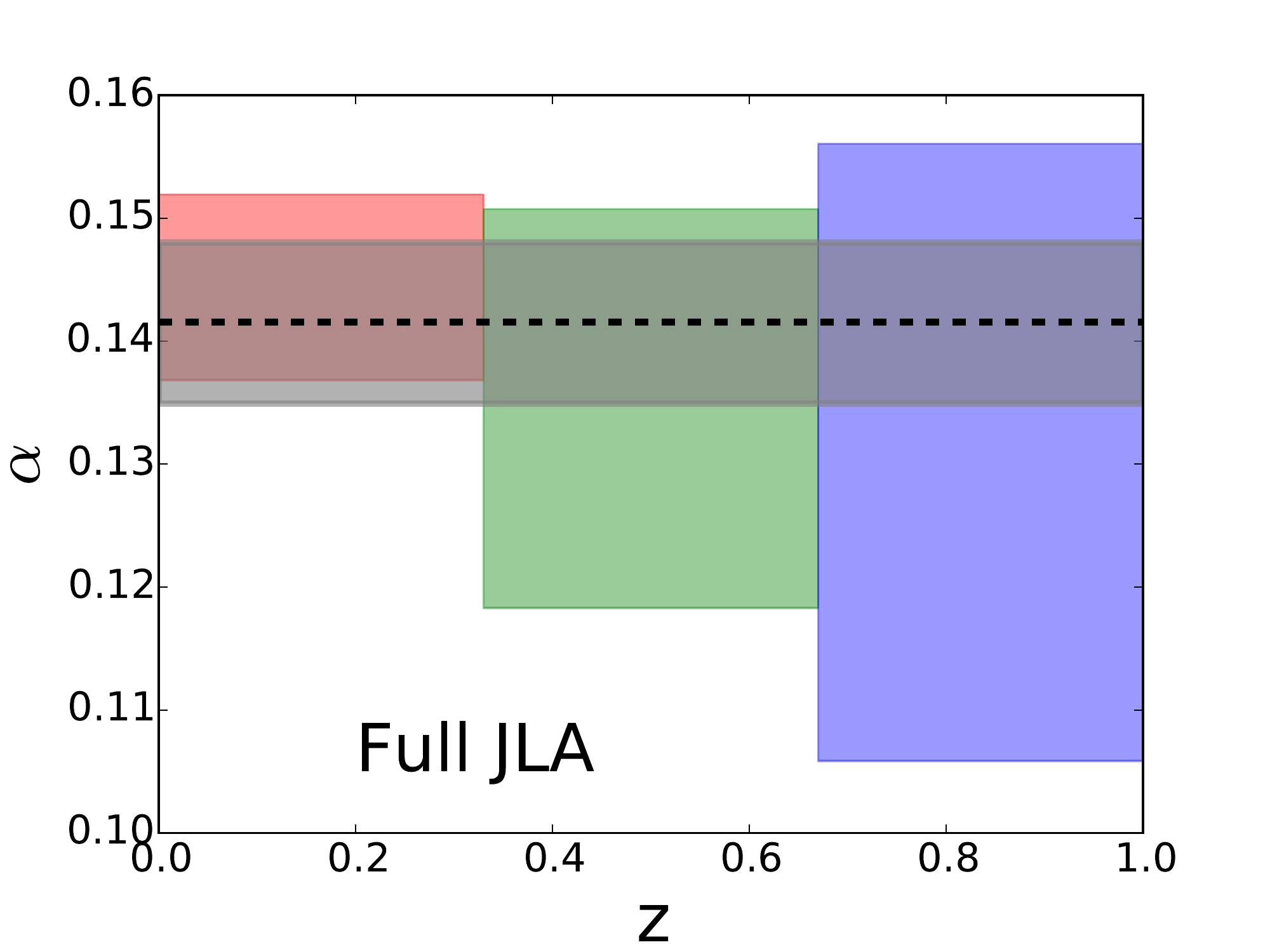}}
      \label{subfig:3bin_alpha_JLA}
      \hspace{0.1\columnwidth}
      \resizebox{0.9\columnwidth}{!}{\includegraphics{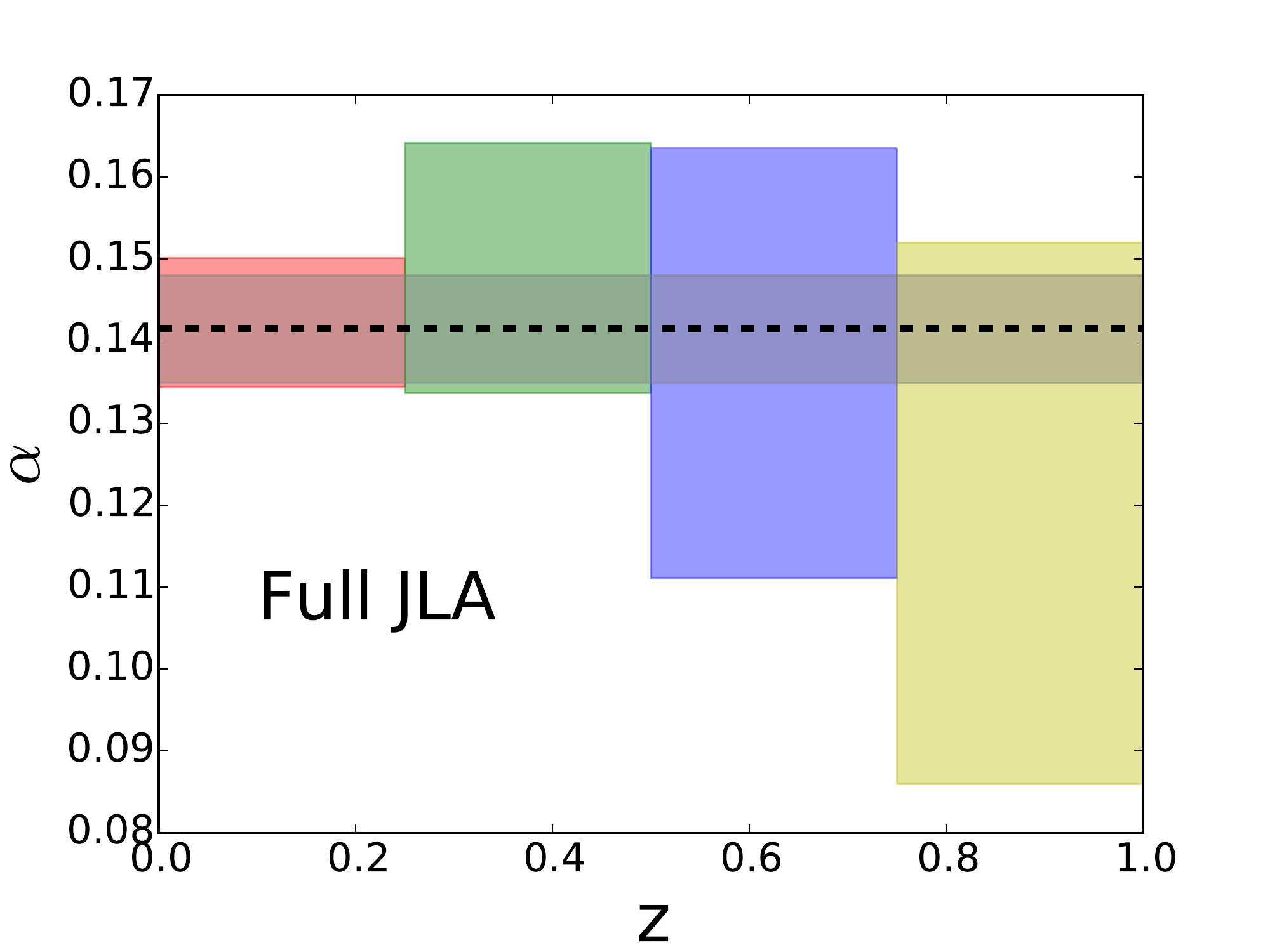}}
      \label{subfig:4bin_alpha_JLA}
      \\
      \hspace{0.1\columnwidth}
      \resizebox{0.9\columnwidth}{!}{\includegraphics{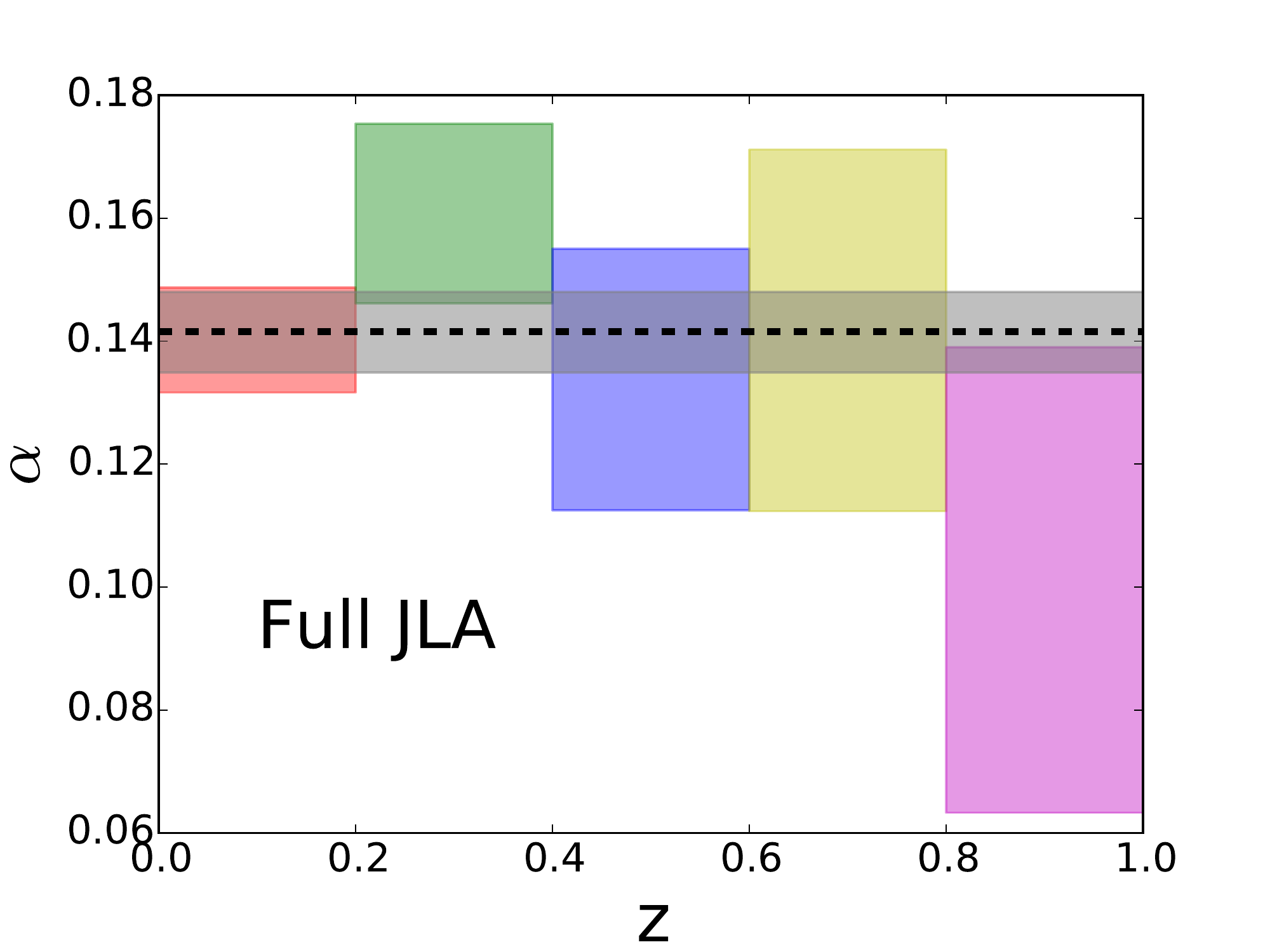}}
      \label{subfig:5bin_alpha_JLA}

    \caption{The 1$\sigma$ confidence regions of stretch-luminosity parameter
      $\alpha$ given by the full JLA sample at redshift region [0,1]. The
      results of 3 bins, 4 bins and 5 bins are shown in the upper left panel,
    the upper right panel and the lower panel. The gray region and the gray dashed line
    denote the 1$\sigma$ region and the best-fit result given by the full JLA
    data. The red, the green, the blue, the yellow and the purple regions
    correspond to the 1$\sigma$ regions of the first, the second, the third,
    the fourth and the fifth bin, respectively.}
               \label{fig:alpha}
\end{figure*}

In Fig \ref{fig:alpha}, we plot the 1$\sigma$ confidence regions of $\alpha$
given by the full JLA sample. The results of 3 bins, 4 bins and 5 bins are
shown in the upper left panel, the upper right panel and the lower panel of
Fig~\ref{fig:alpha}, respectively. For all the panels, it can be seen that
the 1$\sigma$ regions of $\alpha$ given by the full JLA sample (gray region)
overlap with the results given by the SN samples of various bins at 1$\sigma$ CL.
So we can conclude that $\alpha$ is consistent with a constant.
Since this conclusion holds true for all the cases of 3 bins, 4 bins and 5 bins.
we can conclude that it is insensitive to the details of redshift tomography.
This conclusion is consistent with the results of previous studies
\cite{Marriner11,Mohlabeng13,WangWang2013,Shariff15}.

\begin{figure*}
    \centering
      \resizebox{0.9\columnwidth}{!}{\includegraphics{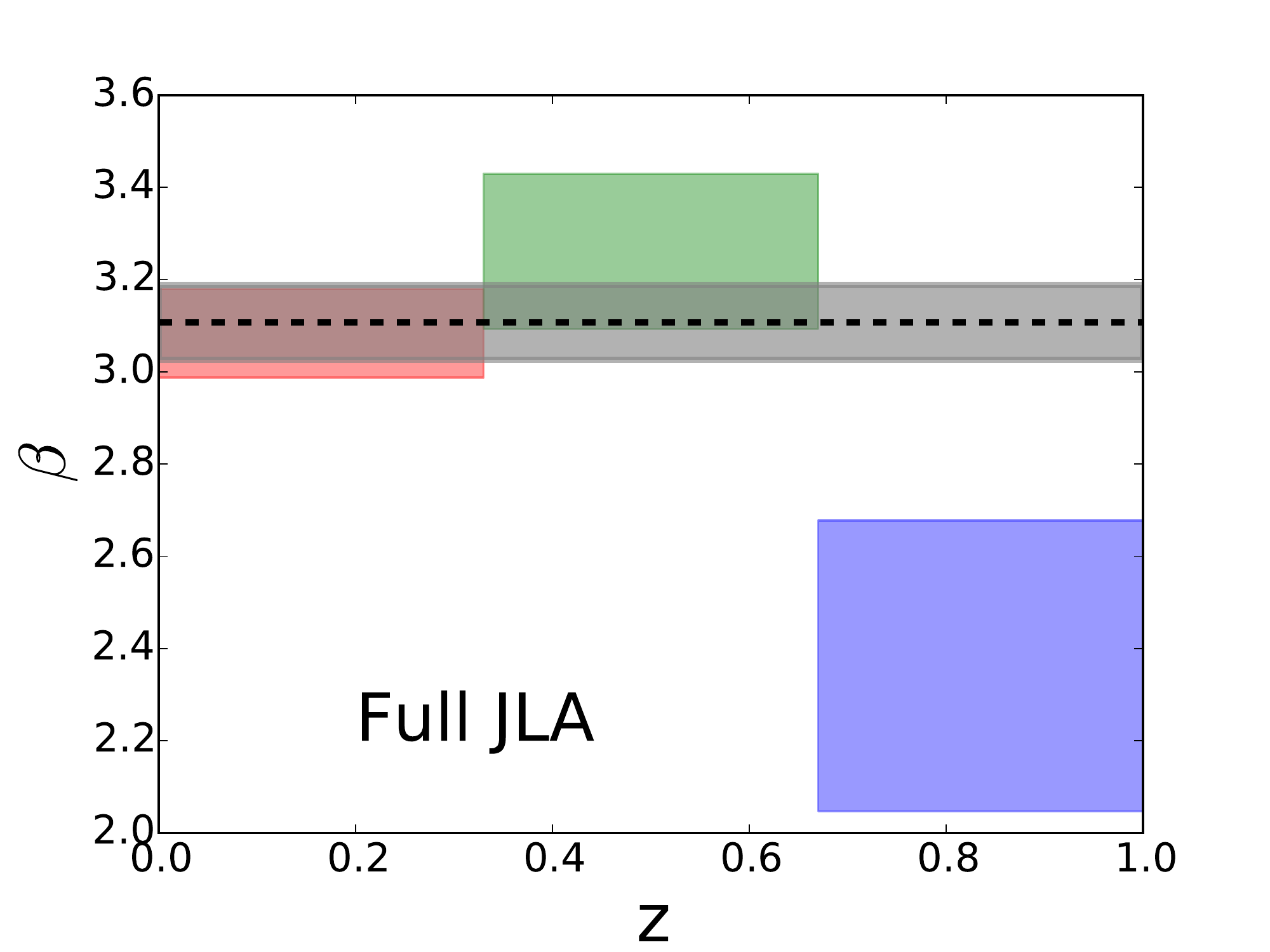}}
      \hspace{0.1\columnwidth}
      \resizebox{0.9\columnwidth}{!}{\includegraphics{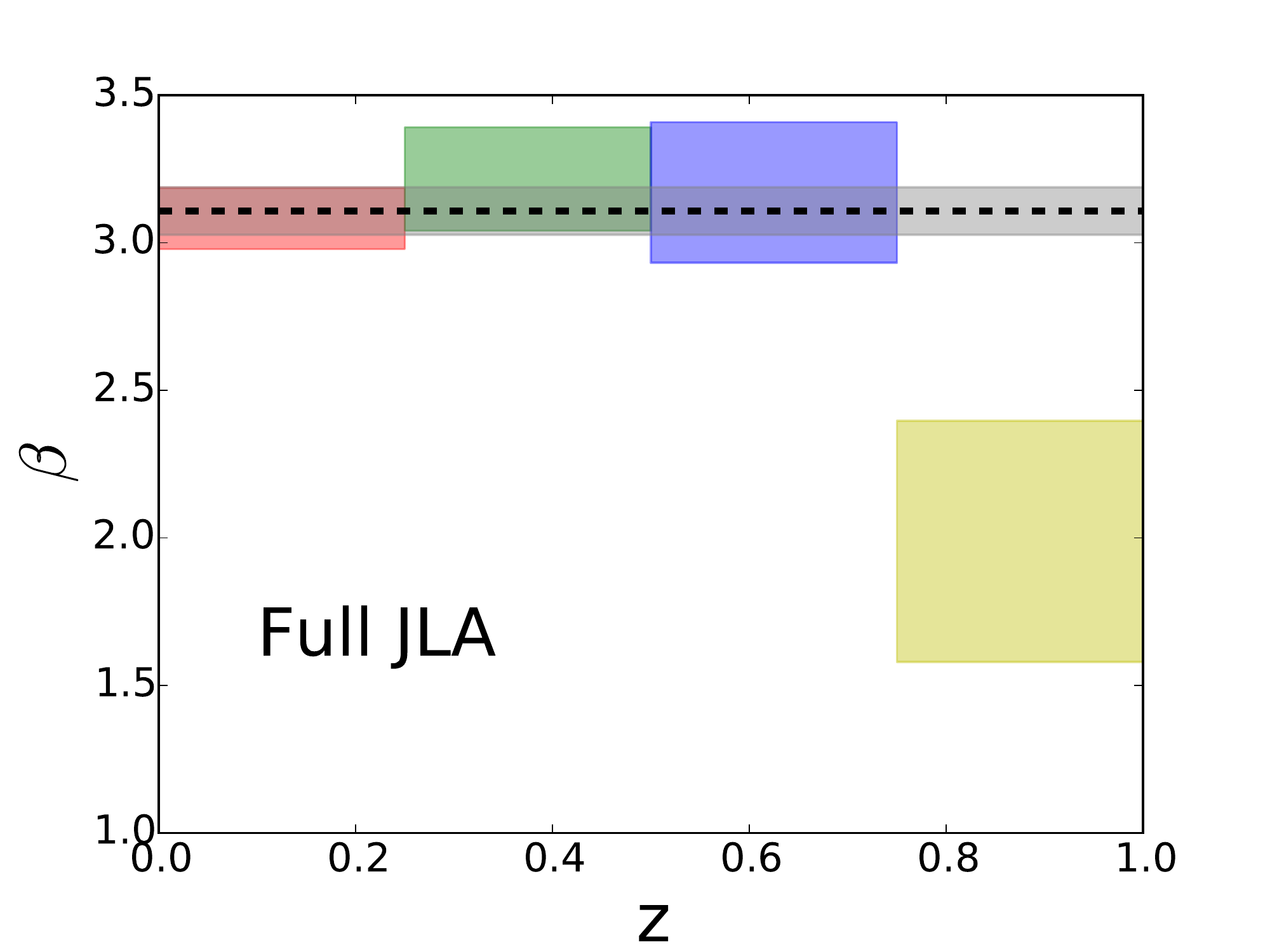}}
      \\
      \hspace{0.1\columnwidth}
      \resizebox{0.9\columnwidth}{!}{\includegraphics{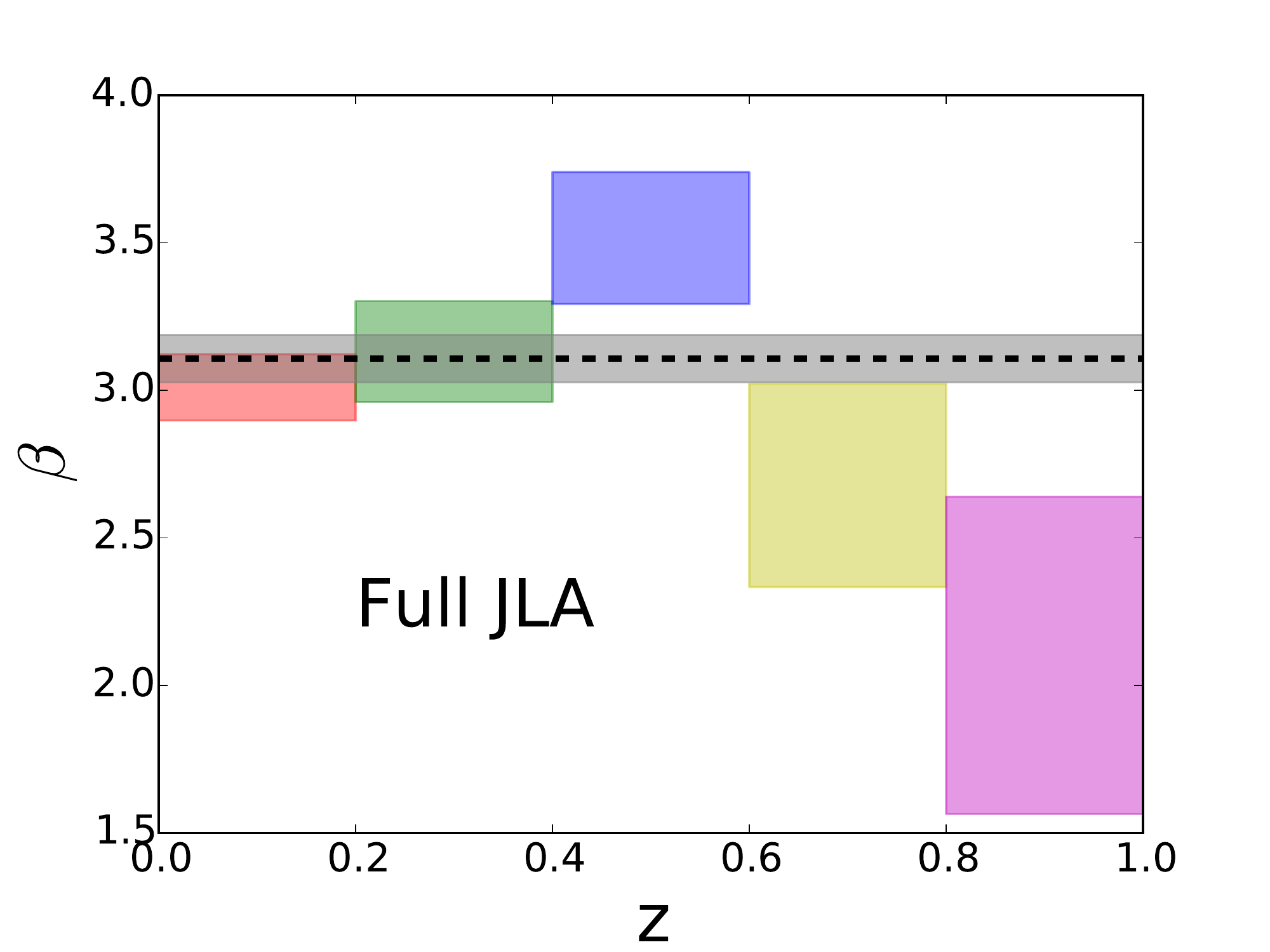}}

    \caption{The 1$\sigma$ confidence regions of color-luminosity parameter
      $\beta$ given by JLA full sample at redshift region [0,1]. The
      results of 3 bins, 4 bins and 5 bins are shown in the upper left panel,
    the upper right panel and the lower panel. The gray region and the gray dashed line
    denote the 1$\sigma$ region and the best-fit result given by the full JLA
    data. The red, the green, the blue, the yellow and the purple regions
    correspond to the 1$\sigma$ regions of the first, the second, the
    third, the fourth and the fifth bin, respectively.}
               \label{fig:beta}
\end{figure*}

In Fig \ref{fig:beta}, we plot the 1$\sigma$ confidence regions of $\beta$ given by the full JLA sample.
The results of 3 bins, 4 bins and 5 bins are shown in the upper left panel,
the upper right panel and the lower panel of Fig~\ref{fig:beta}, respectively.
It can be seen that, although $\beta$ is consistent with a constant at low redshift,
it has a significant trend of decreasing at high redshift.
For the case of 3 bins, the 1$\sigma$ upper bound of $\beta$ in the last bin
deviates from the results given by the full JLA sample at 3.5$\sigma$ CL.
For the case of 4 bins, the 1$\sigma$ upper bound of $\beta$ in the last bin
deviates from the results given by the full JLA sample at 3.6$\sigma$ CL.
For the case of 5 bins, there is a hint for the evolution of $\beta$ for the
fourth bin; moreover, the 1$\sigma$ upper bound of $\beta$ in the last bin
deviates from the results given by the full JLA sample at 3.6$\sigma$ CL.
These results indicate that there is a $\sim$
$3.5\sigma$ CL evidence for the decrease of $\beta$ at high redshift, which is
insensitive to the details of redshift tomography \footnote{To further confirm this point,
we move all the bins $1/4$ bin width to the right and $1/4$ bin width
to the left; then we check whether or not there are any significant differences
for these two cases. It is found that moving bins in such a way will not yield any
significant changes. Therefore, we conclude that the conclusion of $\beta$'s
evolution is insensitive to the details of redshift tomography.}.
It must be stressed that, this conclusion is consistent with the results of some other SN samples \cite{Marriner11,Mohlabeng13},
but is inconsistent with the results of the SNLS3 dataset,
which indicates that $\beta$ has a trend of increasing at high redshift \cite{WangWang2013}.
The reason of this tension is still unclear and deserves further studies.

As mentioned above, JLA dataset includes 118 SNe at $0<z<0.1$ from the low-$z$,
374 SNe at $0.03<z<0.4$ from the SDSS,
239 SNe at $0.1<z<1.1$ from the SNLS,
and 9 SNe at $0.8<z<1.3$ from HST.
It is interesting to explore the evolution of $\beta$ in various
subsamples of JLA.
In this paper we only directly apply the redshift tomography method to
the low-$z$, the SDSS, and the SNLS subsamples, because the HST subsample only contains 9 data points.
To study the effects of HST subsample,
we compare the results of the full JLA sample with the results of the ``JLA without HST'' data.

\begin{figure*}
    \centering
      \resizebox{0.9\columnwidth}{!}{\includegraphics{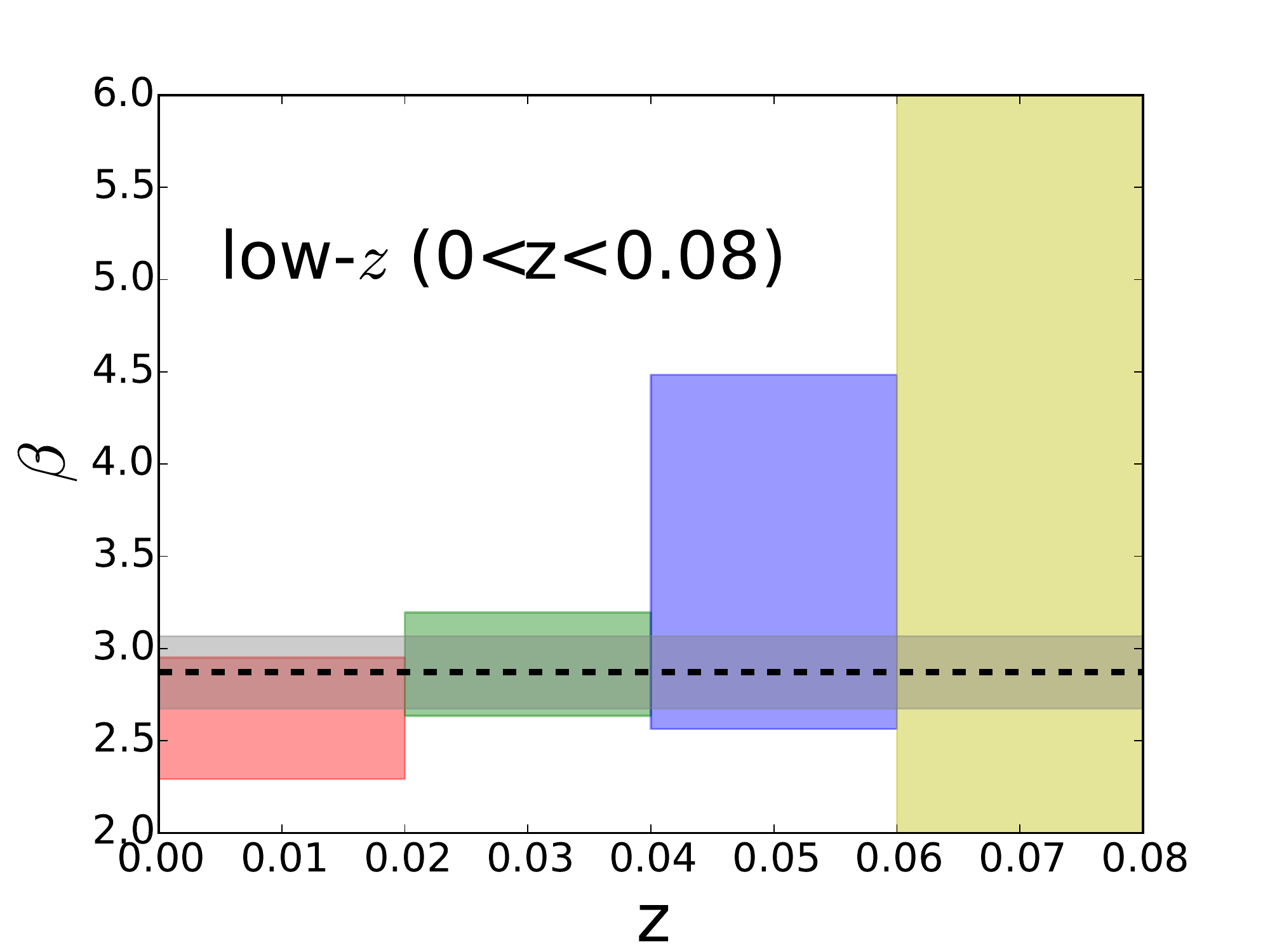}}
      \hspace{0.1\columnwidth}
      \resizebox{0.9\columnwidth}{!}{\includegraphics{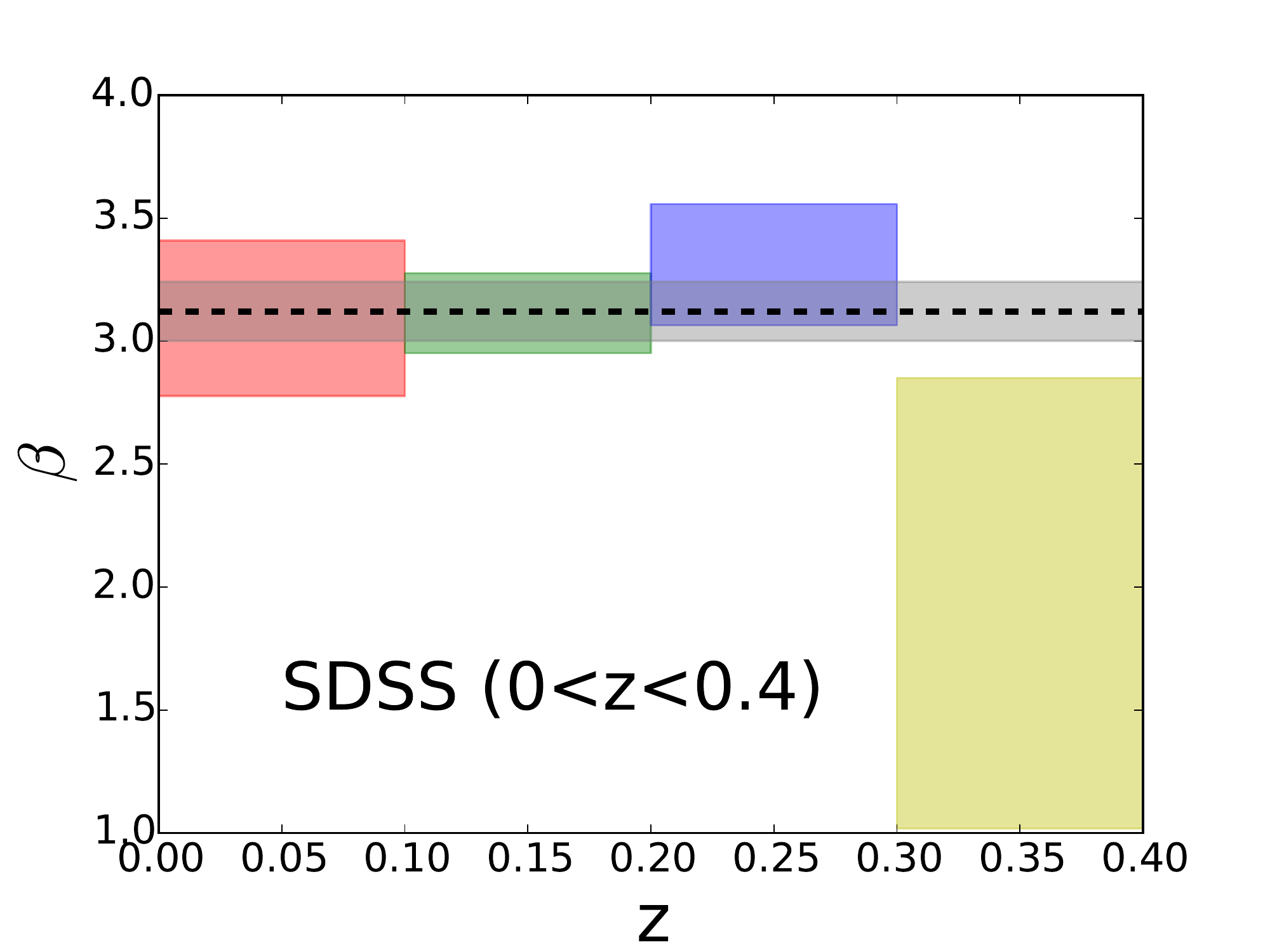}}
      \\
      \hspace{0.1\columnwidth}
      \resizebox{0.9\columnwidth}{!}{\includegraphics{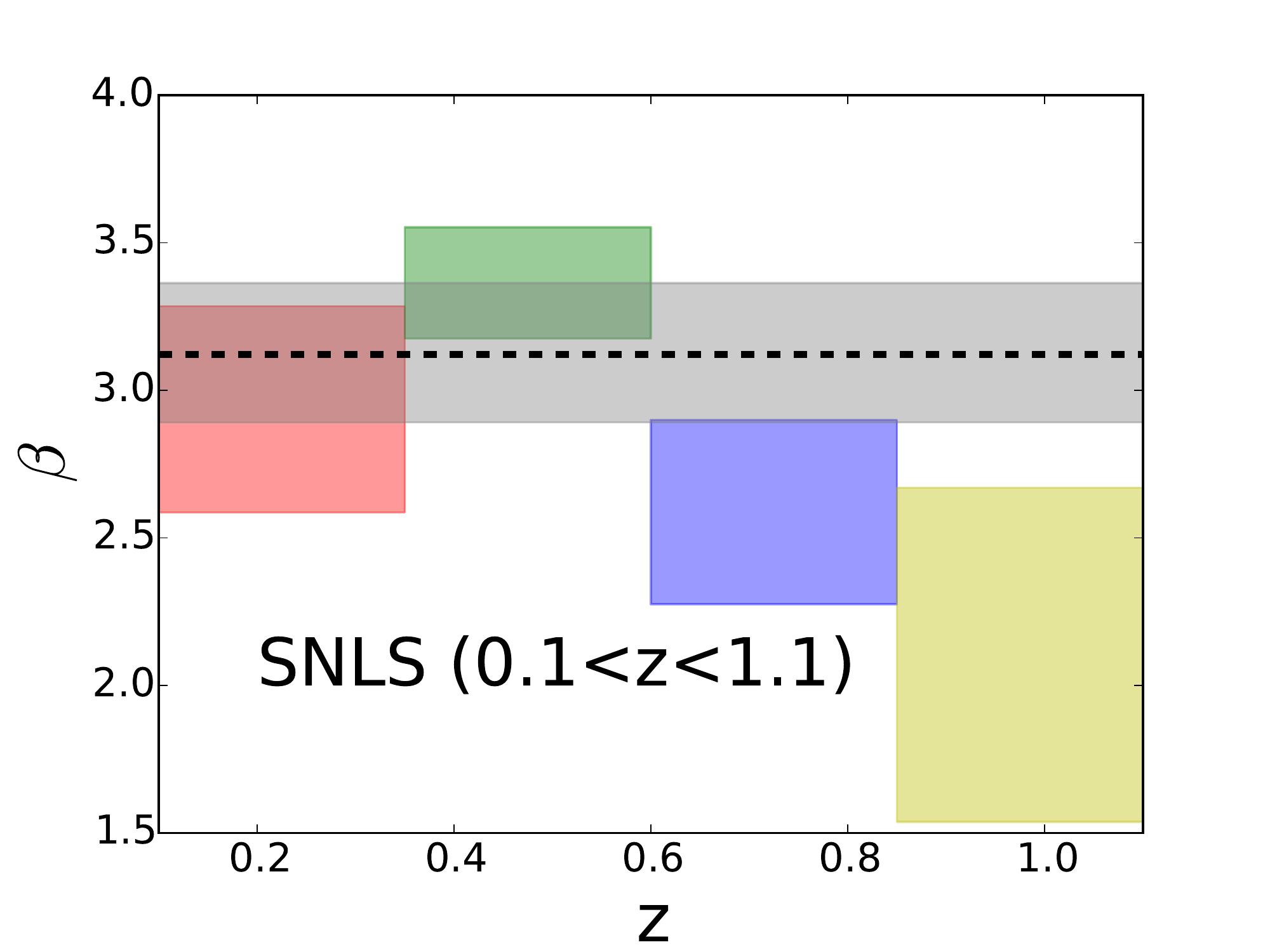}}

      \caption{The 1$\sigma$ confidence regions of
        $\beta$ given by the three subsamples: low-$z$ (upper left panel), SDSS
        (upper right panel) and SNLS (lower panel). The gray region and the gray
        dashed line are the 1$\sigma$ region and the best-fit result given by
        the full low-$z$, the full SDSS and the full SNLS subsample, respectively.
        The red, the green, the blue and the yellow regions correspond to the 1$\sigma$ regions of
        the first, the second, the third and the fourth bin, respectively.}
               \label{fig:subsamples}
\end{figure*}

In the Fig~\ref{fig:subsamples}, making use of the redshift tomography method,
we show the 1$\sigma$ confidence regions of $\beta$ given by each subsample.
The results given by the low-$z$, the SDSS, and the SNLS subsamples
are shown in the upper left panel, the upper right panel, and the lower panel of Fig~\ref{fig:subsamples}, respectively.
For simplicity, here we only consider the case of 4 bins.
For the case of low-$z$, $\beta$ is always consistent with a constant.
For the case of SDSS, the 1$\sigma$ upper bound of $\beta$ in the last bin
deviates from the results given by the full SDSS subsample at 2$\sigma$ CL,
showing that the SDSS subsample favors a decreasing $\beta$ at high redshift.
This conclusion is consistent with the results of \cite{Marriner11}.
For the case of SNLS, the 1$\sigma$ upper bounds of $\beta$ in the third bin and the fourth bin
deviate from the results of the full SNLS subsample at 1.6$\sigma$ and 3.3$\sigma$ CL, respectively.
So compared with the case of SDSS, the SNLS subsample favors a time-varying $\beta$ with a larger decreasing rate.
It should be mentioned that, this conclusion is different from the results of the full SNLS3 sample \cite{WangWang2013}.
This means that SNLS3 dataset may exist some unknown systematic uncertainties \cite{Betoule2014}.

\begin{figure*}

      \centering
      \resizebox{0.9\columnwidth}{!}{\includegraphics{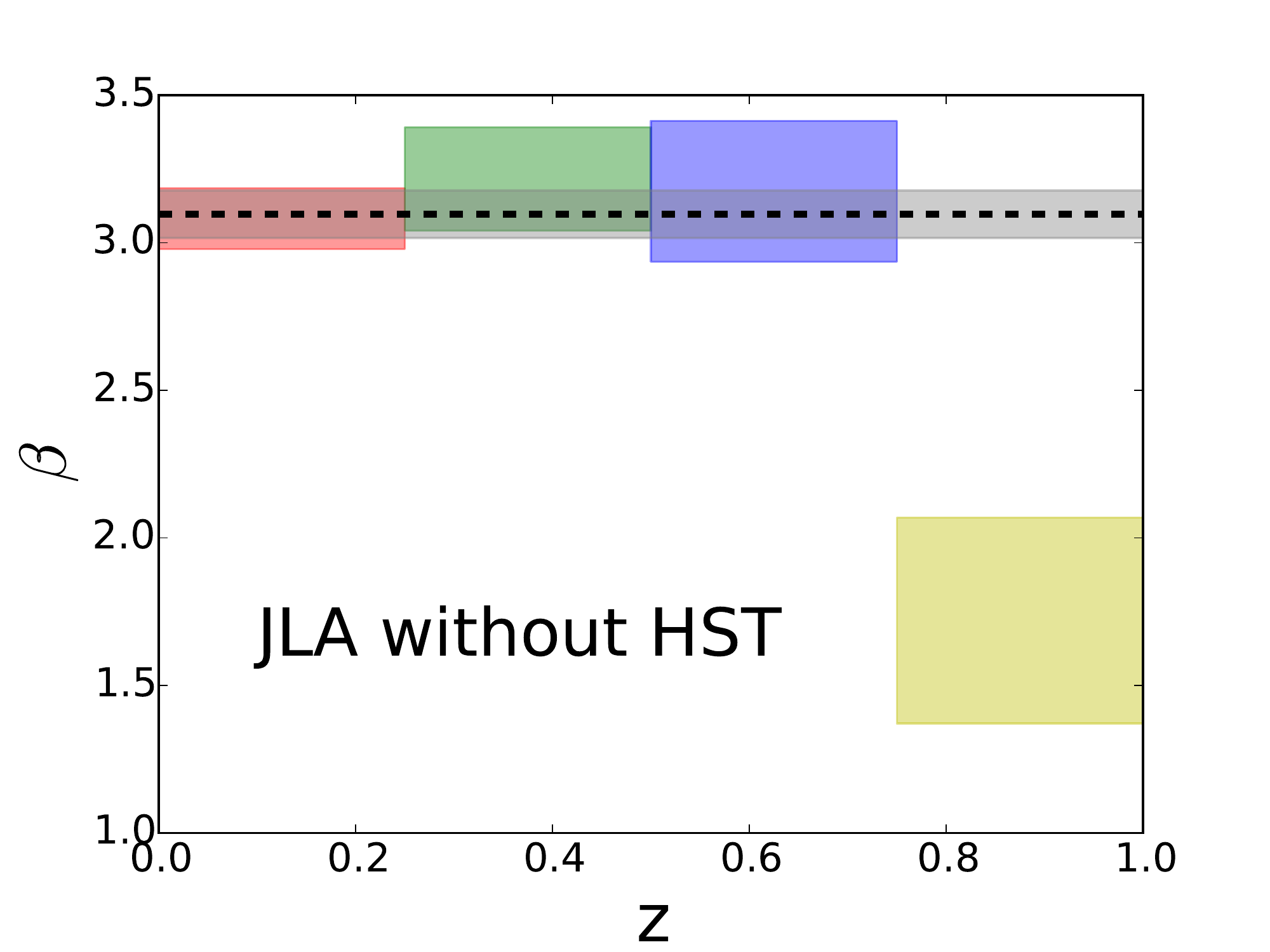}}
      \hspace{0.1\columnwidth}
      \resizebox{0.9\columnwidth}{!}{\includegraphics{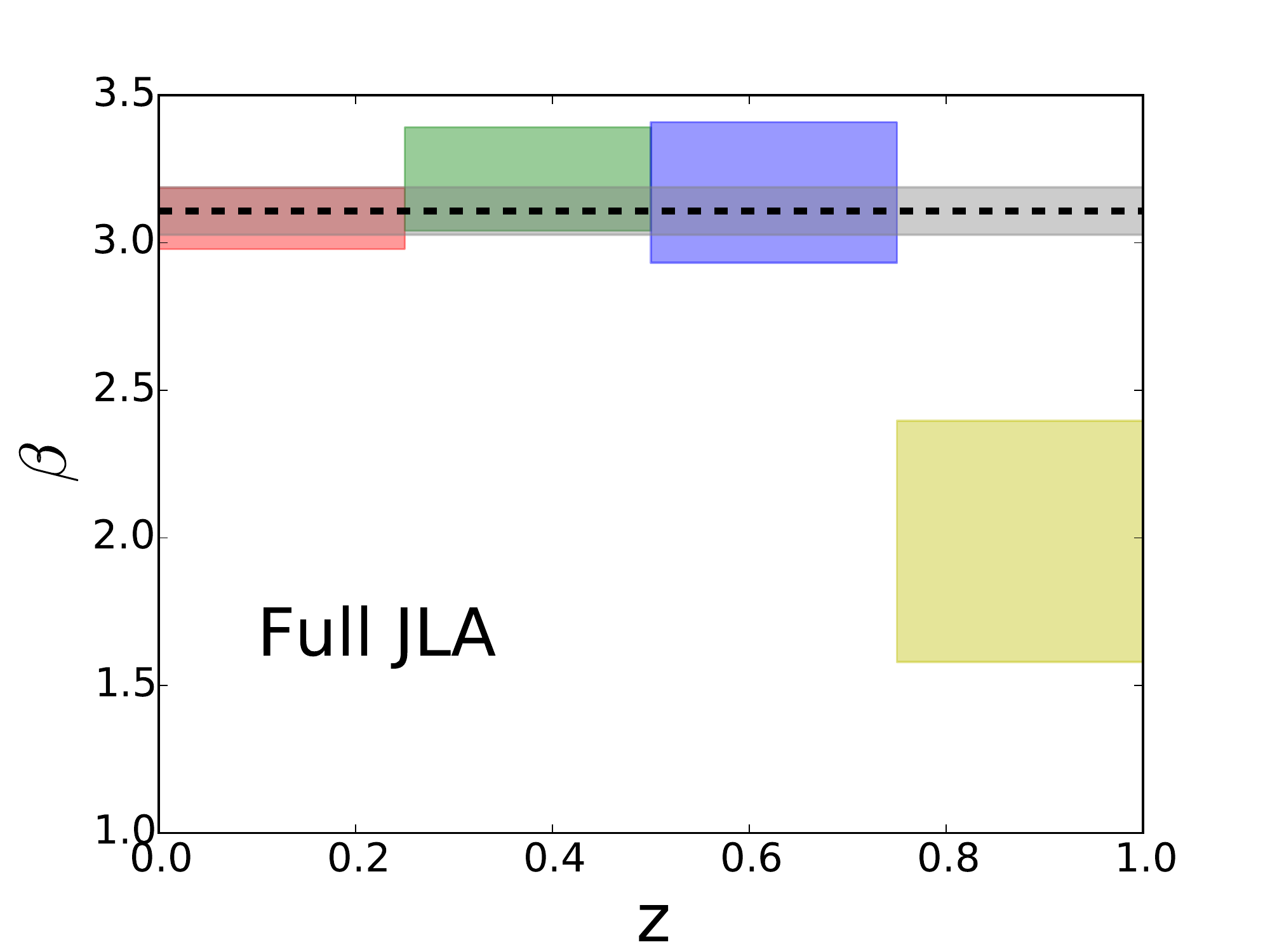}}

      \caption{The 1$\sigma$ confidence regions of
        $\beta$ given by the ``JLA without HST'' data (left panel) and the full JLA sample (right panel)
        at redshift region [0,1]. The gray regions and the gray dashed lines denote
        the 1$\sigma$ regions and the best-fit results given by the full
        samples, respectively. The red, the green, the blue and the yellow
        regions correspond to the 1$\sigma$ regions of the first, the second, the
        third, and the fourth bin, respectively.}

        \label{fig:HST}

\end{figure*}

In Fig~\ref{fig:HST}, we compare the 1$\sigma$ confidence regions of $\beta$ given by
the ``JLA without HST'' data (left panel)
with the results given by the full JLA sample (right panel).
We can see that the HST subsample only affects the evolution behavior of $\beta$ at high redshift.
For the case without HST, the 1$\sigma$ upper bound of $\beta$ in the last bin
deviates from the results given by the full sample at 3.9$\sigma$ CL.
For the case of full JLA sample, the 1$\sigma$ upper bound of $\beta$ in the last bin
deviates from the results given by the full JLA sample at 3.6$\sigma$ CL.
This indicates that HST subsample can slightly slow down the decreasing rate of $\beta$ at high redshift.

\begin{figure}
      \centering
      \resizebox{0.9\columnwidth}{!}{\includegraphics{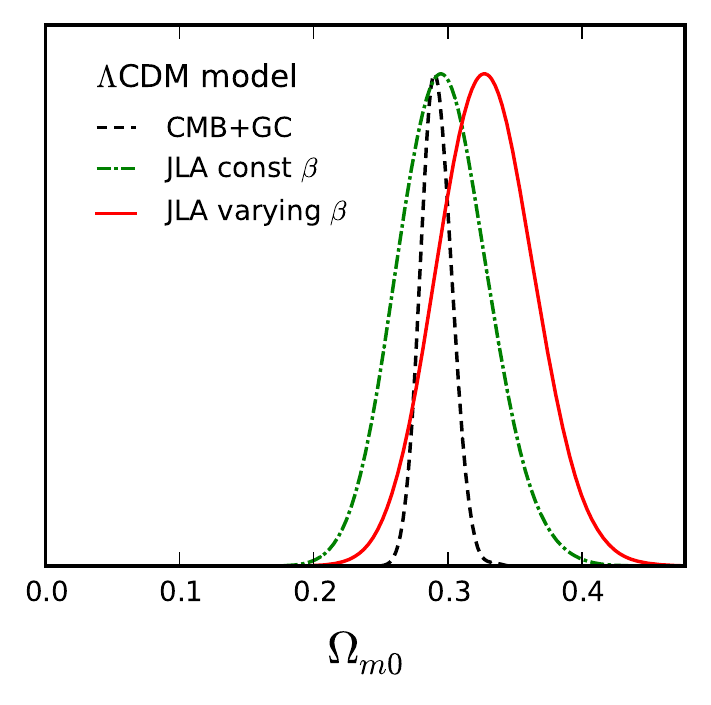}}

      \caption{The 1D marginalized probability distributions of $\Omega_{m0}$
        given by the full JLA sample for the $\Lambda$CDM model. Both the results
        of constant $\beta$ (green dash-dotted line) and varying $\beta$ (red
        solid line) cases are presented. The corresponding results given by the
      CMB+BAO data (black dashed line) are also shown for comparison.}

        \label{fig:Om0}
\end{figure}

Next, we discuss the impacts of a varying $\beta$ on the parameter
estimation. For simplicity, here we consider the standard cosmological model:
the $\Lambda$CDM model. As shown in Fig \ref{fig:beta}, $\beta$ perfers a higher value at low
redshift and a lower value at high redshift. So we assume that $\beta$ is related
to the redshift by a simple piecewise function
\begin{equation}
  \label{eq:piecewisebeta}
  \beta(z) = \left \lbrace
   \begin{array}{ll}
    \beta_1 &\quad   0<z \leq 0.75\,,\\
    \beta_2 &\quad 0.75<z
    \end{array}
    \right.
\end{equation}
where $\beta_1$ and $\beta_2$ are two model parameters. In Fig \ref{fig:Om0}, by using
the full JLA sample only, we plot the 1D marginalized probability distributions of $\Omega_{m0}$ in
the cases of constant $\beta$ and varying $\beta(z)$. It can be seen that varying
$\beta$ yields a lager $\Omega_{m0}$ than the case of constant $\beta$: for the
case of varying $\beta$, the best-fit value of $\Omega_{m0}$ is $0.329$, while for
the case of constant $\beta$, the best-fit value of $\Omega_{m0}$ is $0.297$. Note that our
result is consistent with the results of \cite{Shariff15}. To make a
comparison, in Fig \ref{fig:Om0} we also plot the 1D marginalized probability distribution of
$\Omega_{m0}$ given by a combination of the CMB \cite{Planck2015}
\footnote{In addition to \cite{Planck2015}, there are some other
distance priors data, e.g. see Refs. \cite{WangDai15,Huang2015,WW13}.}
and the Baryon Acoustic Oscillations (BAO)
\cite{Hemantha14,Wang2014} data. The best-fit value of
$\Omega_{m0}$ given by CMB+BAO data is $0.292$, which is closer to the
best-fit value of the constant $\beta$ case. This result is different from the
result of the SNLS3 sample \cite{WLZ14}. However, the result of $\Omega_{m0}$ for the varying $\beta$ case
is still consistent with the result for the constant $\beta$ case,
as well as the result given by the CMB+BAO data, at $1\sigma$ CL.

\section{Summary}
\label{sec:conclusion}
SN Ia is one of the most powerful tools to explore the current cosmic acceleration.
As the sample size of SN Ia rapidly grows, it is very important to perform
various tests for these SN samples. One of the most interesting tests is
to probe the possible evolution of SN color parameter, which has
drawn a lot of attentions in recent years
\cite{Marriner11,Mohlabeng13,WangWang2013,Shariff15}.
\footnote{In addition to exploring the possible evolution of $\beta$, there are
some other meaningful tests for SN smaples, e.g. see Refs.
\cite{Bengochea11,Kim11,Hu15,WCM12,Wang2000}}

In a latest work \cite{Shariff15}, adopting two particular
parameterizations of $\beta$, Shariff et al.
found 4.6$\sigma$ CL evidence for a significant drop in $\beta$ at redshift
$z=0.66$, for the JLA sample. In the current work,
we revisit the possibility of $\beta$'s evolution by using the redshift tomography method.
In addition to the full JLA sample, we also study the cases of various JLA subsamples.
So far as we know, the effects of various JLA subsamples on $\beta$'s evolution
have not been studied in the past. Moreover, we also briefly discuss the impacts of time-varying
$\beta$ on parameter estimation are also studied.

Our conclusions are as follows:
\begin{itemize}
\item
If the full JLA sample is used, then
$\alpha$ is always consistent with a constant
(see Fig~\ref{fig:alpha}), and $\beta$ has a significant
trend of decreasing, $\sim 3.5\sigma$ CL, at high redshift
(see Fig~\ref{fig:beta}). It should be pointed out that, due to that the redshift tomography
method tends to reduce statistical significance, the redshift-dependence
of $\beta$ is studied the hard way in this work. Since the effect of $\beta$'s
evolution is strong enough to be found after adopting the redshift tomography method,
we can conclude that the evolution of $\beta$ is indisputable.

\item
If the low-$z$ subsample of JLA is used,
then a constant $\beta$ is favored. In contrast, if the SDSS or the SNLS subsamples is
adopted, then a decreasing $\beta$ is favored. Besides,
compared with SDSS subsample, SNLS subsample prefers a larger decreasing rate of $\beta$
(see Fig~\ref{fig:subsamples}). It should be pointed out that the trajectory
of $\beta$ given by the SNLS subsample of JLA is quite different from the
prediction of the full SNLS3 sample \cite{WangWang2013}. This means that the SNLS3
dataset may have some unknown systematic bias, or anomalies, not accounted for by
the reported systematic uncertainties of SNLS3.
\item
If the HST subsample is removed from the full JLA data,
then the decreasing rate of $\beta$ at high redshift will be slightly enlarged
(see Fig~\ref{fig:HST}).
\item
If a binned parameterization of $\beta$ is adopted, then a larger best-fit value of
$\Omega_{m0}$ will be obtained, compared to the
case of constant $\beta$. However, if the information of $1\sigma$ region is taken into
account, then for both the time-varying $\beta$ and the constant $\beta$ cases,
the results of $\Omega_{m0}$ are consistent with the result given by the
CMB+BAO data.
\end{itemize}

In this paper, we only consider the simplest $\Lambda$CDM model.
In addition to $\Lambda$CDM,
many other DE models \cite{Li2004,Chevallier01,Linder03} are also favored by current cosmological observations.
It is of interest to study the effects of varying $\beta$ on parameter estimation
in other dark energy models~\cite{Zlatev1999,Caldwell2002,Li2004,WZ2008,WZX2008,
LiLiWang2009a,LiLiWang2009b,Huang2009,LLLW10,WLL10,WLL11,LLWWHM11,ZLLWZ12,LWLZ13,
HLLW15,WHML16}. This will be done in future works.

\section*{Acknowledgments}

We are very grateful to the referee for the valuable suggestions.
We also thank Prof. Yi Wang for carefully reading the manuscript of this work.
ML is supported by the National Natural Science Foundation of China (Grant No. 11275247, and Grant No. 11335012)
and 985 grant at Sun Yat-Sen University.
SW is supported by the National Natural Science Foundation of China under Grant No. 11405024
and the Fundamental Research Funds for the Central Universities under Grant No. 16lgpy50.



\label{lastpage}

\end{document}